\documentclass[aps,pre,twocolumn,groupedaddress]{revtex4-1}
\usepackage{graphicx}
\usepackage{comment}
\usepackage{siunitx}
\usepackage{gensymb}
\usepackage{amsmath,amsfonts,amssymb,amsbsy}
\usepackage[hidelinks]{hyperref}
\newcommand\Tstrut{\rule{0pt}{2.6ex}}         

\usepackage{color}
\usepackage[colorinlistoftodos]{todonotes}
\usepackage{verbatim}

\begin{document}

\title{Friction controls submerged granular flows}

\author{\underline{Juha Koivisto}$^{1,2}$}
\author{Marko Korhonen$^{1}$}
\author{Mikko Alava$^{1}$} 
\author{Carlos P. Ortiz$^{2}$}
\author{Douglas J. Durian$^{2}$}
\author{Antti Puisto$^{1}$}
\affiliation{\\$^{1}$Department of Applied Physics, Aalto University, Aalto 00067, Finland}
\affiliation{$^{2}$Department of Physics and Astronomy, University of Pennsylvania, Philadelphia,\\ Pennsylvania 19104-6396, USA\\}

\date{\today}
\begin{abstract}
We investigate the coupling between interstitial medium and granular 
particles by studying the hopper flow of dry and submerged system 
experimentally and numerically.
In accordance with earlier studies, we find, that the dry hopper 
empties at a constant rate. However, in the submerged system we 
observe the surging of the flow rate. 
We model both systems using the discrete element method, which we couple 
with computational fluid dynamics in the case of a submerged hopper. 
We are able to match the simulations and the experiments with good 
accuracy. 
To do that, we fit the particle-particle contact friction for each 
system separately, finding that submerging the hopper changes the
particle-particle contact friction from $\mu_{vacuum}=0.15$ to 
$\mu_{sub}=0.13$, while all the other simulation parameters remain 
the same. Furthermore, our experiments find a particle size dependence 
to the flow rate, which is comprehended based on arguments on the terminal 
velocity and drag. 
These results jointly allow us to conclude that at 
the large particle limit, the interstitial medium does not matter, in 
contrast to small particles. 
The particle size limit, where this occurs depends on the viscosity 
of the interstitial fluid.
\end{abstract}


\maketitle

\section{Introduction}

Understanding the coupling between solid particles and liquid is a challenging task due the complexity of grain-grain and grain-liquid interactions \cite{zhou2010discrete,zhu2007discrete}. Even in vacuum the assemblies of granular particles exhibit highly complex dynamics due to the different possible phases of existence. Depending on the loading and the particle geometry, it can appear in gaseous, fluid-like or solid-like phases \cite{Eshuis2007}. Related to this, the rheological characteristics of granular matter falls into the category of yield stress fluids \cite{Divoux2015,Johnson2017}. However, their behavior is even more complex, as many of them show discontinuous shear thickening at intermediate shear \cite{Seto2013}. Such an effect is attributed to the interparticle friction and/or the interlocking of the grains, depending on their shape \cite{LosertPRE00,Athanassiadis2014, Jaeger2014}.

\begin{figure}[!th]
\includegraphics[width=\columnwidth]{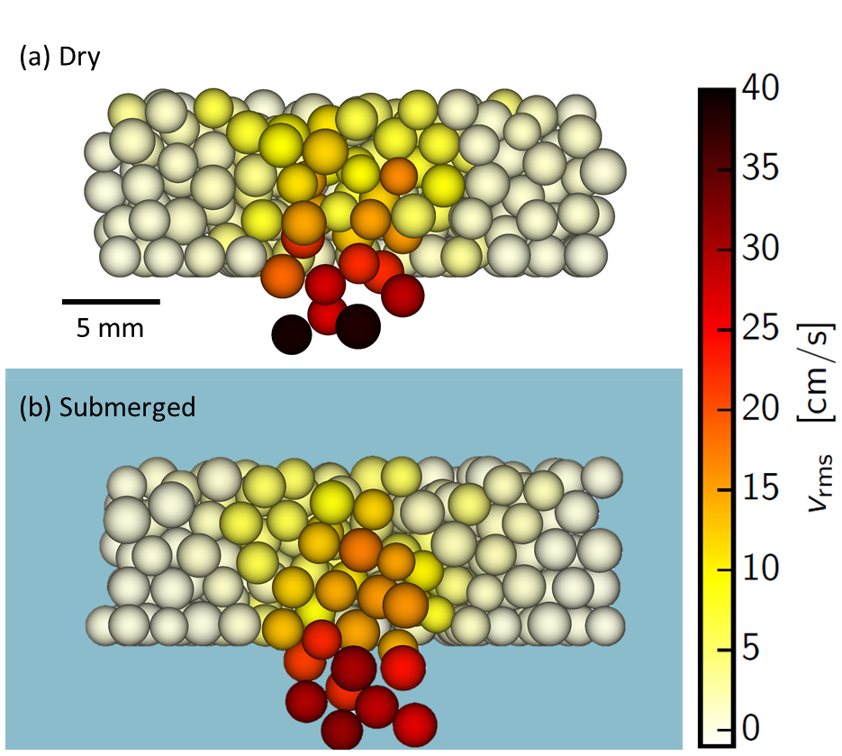}
\caption{\label{fig1}Snapshot of the simulation close from the orifice shows the localized particle velocities. Here, the 3D cylinder is visualized at the center plane. The dry simulation with particles in vacuum (a) has a lower packing fraction than the submerged case with water as interstitial medium (b). In the vacuum system, the flux is fully driven by gravity pulling the particles out from the hopper. In the submerged case in addition to gravity the particle flux is driven by the fluid pushing the particles. Due to this, the particle volume fraction at the exit is higher in the submerged case (b) compared to vacuum (a) as is visually evident.}
\end{figure}

The 3D hopper flow, shown in Fig.~\ref{fig1}, is a well studied model case of grain flow \cite{Thomas2016,Wilson2014,Thomas2015}, partly due to its seeming simplicity, but also due to its importance in practical applications, from simple silos in farms to complex pharmaceutical factories. Even in a simple hopper scenario, all three granular phases exist, and there is the gas phase outside the hopper; near the hopper boundaries the grains are in the bulk or solid phase, while above the orifice, there must be a yielded (fluid) phase enabling the flow.
%

Numerous studies have shown that in a dry hopper, the outflow 
of the granular particles follows the Beverloo equation~\cite{Beverloo1961,Madrid2016Arx}. 
That is, the outward flux of 
grains remains constant in time, until the hopper runs out of grains.
The grain liquid interaction culminates to three components: terminal velocity as well as Darcy and modified Beverloo equations. The Beverloo equation for dry case 
\begin{equation}
W_{dry} = C \rho \sqrt{g}(D-kd)^{(5/2)}
\end{equation}
describes the mass flow rate as a function of density $\rho$, gravity $g$ as well as particle $d$ and orifice $D$ diameters. 
The term $D-kd$ illustrates the empty annulus where the particles partially reduce the size of the orifice through constant $k$. 
The exponent 5/2 can be derived through the so called free fall theorem 
that essentially says that the flow rate is proportional to the particle 
velocity as if it would fall of a dome covering the orifice 
\cite{Tian2015,Janda2012,Mankoc2007}. 
The geometry then dictates the exponent 5/2.

Recently the Beverloo equation is adapted and simplified to the liquid case \cite{Wilson2014} 
\begin{equation}
	W = C\rho v_t d^2 (D/d - k)^2, \label{eq:modified-beverloo}
\end{equation}   
\noindent where the acceleration due gravity in fixed distance is replaced by terminal velocity $v_t$ in a liquid. It is important to note that the flow decreases to zero when the ratio $D/d$ approaches $k$. The $C=0.4$ and $k=2.4$ are the empirical fit parameters in a submerged case \cite{Wilson2014,koivistoSubmitted}.

The Beverloo equation only considers the dimensions of the grains and the 
orifice but not the properties of the interstitial medium, such as viscosity or drag.
Some studies have considered the role of air as an interstitial 
medium \cite{Yuu2011Mat}. There are simulations and experiments that 
show non-trivial flow patterns of air when particles move, starting 
from oscillations \cite{WuX.L.Maloy1993,Bertho2002} and steady state 
turbulent like flows \cite{Hilton2011}. This movement of gas like 
medium affects the flow of grains by creating pressure gradients and 
drag. 
These effects are in this case minor, since the drag caused
by air is rather modest.

When the grains are embedded in a liquid, whose viscosity is orders of magnitude larger compared to that of air,
the effect of interstitial medium is expectedly more pronounced \cite{koivistoSubmitted}. There, the flow rate of grains actually increases in time, i.e.~surges \cite{Wilson2014}. Compared to a Newtonian fluid running out of a bucket this behaves exactly the opposite; there the flow rate decreases as the water runs out. In this submerged granular flow, the complexity of the problem rises from the fluid-particle interactions. As in the dry hopper scenario, the driving force of the system is the particle flow created by gravity. However, here the motion of the particles additionally creates fluid flow that disturbs the particle trajectories. This feedback loop between fluid and particles presumably increases the driving pressure of grains as they run out. A simple analytical model taking this into account is already shown in Ref.~\cite{koivistoSubmitted}.

In this paper we show that one can successfully capture both qualitatively and quantitatively this counter intuitive behavior arising in a submerged granular hopper flow using coupled discrete element model (DEM) for the particle dynamics and computational fluid dynamics (CFD). While the particle trajectories and interactions are computed explicitly in the DEM-implementation, the fluid flow is modeled on a continuum level by the CFD approach. This is fundamentally different from the inertial $\mu(I_v)$-model where the granular media and the interstitial fluid is treated as a single continuum \cite{Boyer11PRL}.

The article is organized as follows: It starts by introducing the reader to our Methods, giving the details of both the experiments and the simulations. Then in the section Results we describe the main findings, showing that the features observed in experiments are captured by the simulations. Once the validity of the simulation is confirmed, the values of grain-grain friction is swept using simulations.
The article finishes with Conclusions, where we discuss the results, and give the readers a short overlook to future research.
 
\section{Methods}

Here, we study both the dry and submerged granular hopper flows.
In the simulations, we assume that we can model the dry case without 
the interstitial fluid (no CFD), since  air viscosity and density are 
negligible.
In contrast, the submerged granular flow comprises two distinct 
phases (granular particles and liquid) that interact by various forces 
and have to be modeled concurrently. 
The approach adopted here is to model the liquid phase on a continuum 
level and the granular phase as discrete particles. Specifically, the 
fluid phase is modeled by the Computational Fluid Dynamics (CFD) 
method~\cite{ferziger2012computational}, which utilizes the Finite Volume 
Method (FVM) for discretizing the Navier-Stokes in the problem domain. 
The Discrete Element Method (DEM)~\cite{cundall1979discrete} is applied 
for the granular (particle) phase and each particle trajectory is 
integrated individually based on the interaction forces.

In the CFD framework, the modified Navier-Stokes equations (NSEs)
~\cite{anderson1967fluid}, physically implying the conservation of mass 
and momentum, are discretized and solved to yield the relevant quantities, 
such as the local fluid velocity and pressure fields. The modified NSEs 
(here, Eqs. (23) and (40) in Ref.~\cite{anderson1967fluid}) include the 
particle-fluid interaction term, that contains the sum of the appropriate 
interaction forces, such as the drag force \cite{DiFelice1994}, buoyancy, 
pressure gradient forces and the imposed shear stress~\cite{zhu2007discrete}. 
This term is also present in the DEM scheme, where it is included in the 
Newton's 2nd law which is formulated and solved for each particle. 
This coupling scheme has the inherent advantage of providing an accurate 
description of both the fluid and the particle phase at a reasonable 
computational expense~\cite{zhu2007discrete}.

The CFD-DEM coupling is realized in a readily implemented software called CFDEM project which combines
the OpenFOAM CFD-library with a DEM solver (LIGGGHTS~\cite{kloss2012models}), providing the user  extensive control over the simulation particulars and more importantly, the NSEs and fluid-particle interaction models~\cite{zhou2010discrete}. The implementation also grants efficient CPU parallel execution via the Message Passing Interface (MPI).

The material parameters used in the numerical method are obtained, where possible, from the experiments or utilizing textbook values.
In the experiments, there are three types of grains, while in the simulations only the largest one is used. The grains are technical quality soda lime silica glass beads with $d=0.2\pm0.01$ cm (A-205),  $d=0.1\pm0.01$ (A-100) and $d=0.05\pm0.005$ (P-230) in diameter from Potters Industries. Their density is $\rho=2.54 \pm 0.01~\mathrm{g/cm^3}$ measured using the Archimedes method by sinking the beads in liquid and measuring the weight of the grains and fluid volume displacement.
The simulations additionally require knowledge of the elastic (Young's and shear) moduli, and the friction and restitution coefficients.
The typical values for Young's and shear moduli of glass beads tabulated in textbooks are $E=72$ GPa and $G=30$ GPa, respectively.
These give the Poisson's ratio of $p = E/(2G) - 1 = 0.2$.
These values of the grain properties were set in the simulations to match the experimental values.

The friction and restitution coefficients, describing the dissipation of the grain-grain contacts and collisions, are the 
remaining parameters required to perform DEM simulations. Measurement of either of these for glass beads is impractical as it requires to estimate the dissipated energy in a dense granular flow. A textbook value for sliding of wet glass surfaces is around $\mu=0.1$, which can be taken as a starting point for the simulations. 
A sensible value for the restitution coefficient of hard-sphere-like glass beads is $\alpha=0.9$. For instance, in similar dry simulations involving softer grains, the restitution coefficient of $\alpha=0.8$ has been used \cite{SchwartzGM12}. 


In the setup, the liquid phase consists of filtered tap water at $T=22~\degree C$ temperature with the well known textbook values 
for viscosity $\eta = 1.0$~mPa$\cdot$s and density $\rho_f=1.00~\mathrm{g/cm^3}$. Accordingly, these values were used in the 
simulations, with the further assumption of laminar flow conditions. 
Laminar flow can be safely assumed owing to the fact 
that the flow rates remain rather modest being purely 
driven by the release of the grains' potential energy.
In practice the hopper is submerged in a large fish tank. There are 
no water -- air interfaces. The experiment is totally under water. 
The scale is above the water level, measuring the weight of 
the remaining beads in the hopper. 
The hopper is a flat 
bottomed cylindrical tube made of transparent polycarbonate with $D_h = 5.0$~cm diameter. 
The orifice is a circular hole ($D=1.0$~cm) with 1~mm vertical walls that expand in 45-degree bevel cut at the center of the aluminum bottom. 
The experimental setup is described in detail in Refs.~\cite{koivistoSubmitted,koivisto2017PRE} and their supplemental material.



\begin{figure}
\includegraphics[width=\columnwidth]{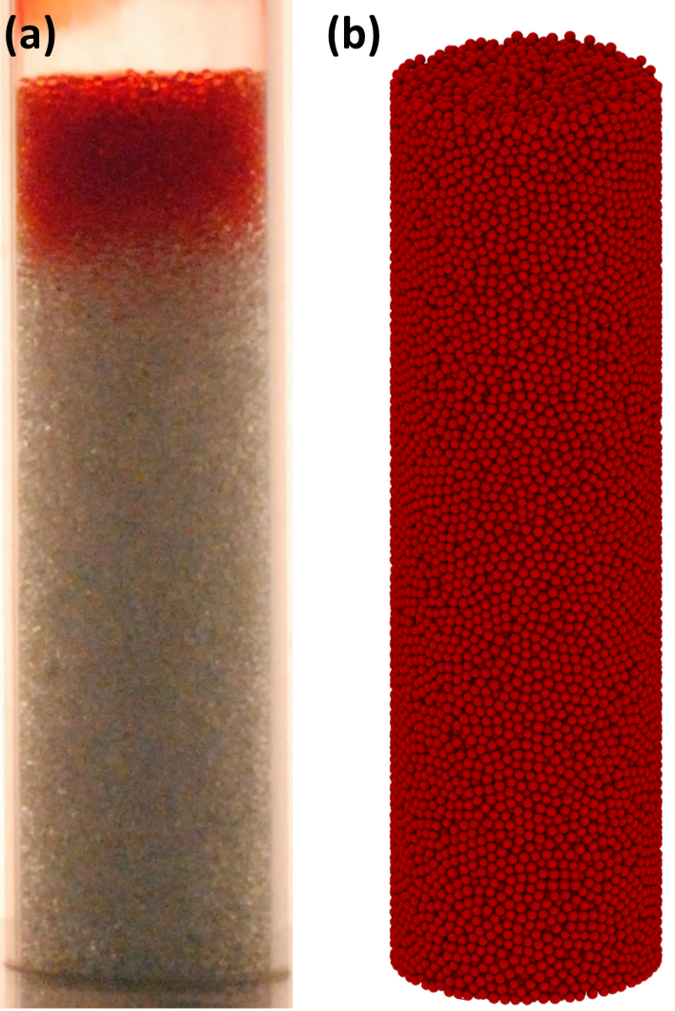}
\caption{Experimental (a) and numerical (b) geometries are the same; The orifice diameter is $D=1.0$ cm, particle diameter $d=0.2$ cm and hopper diameter $D_h=5.0$ cm. The red dye injected on top of the grains in the experiments (a) is visualizing the fluid flow. The grains are transparent glass but appear albescent.
\label{fig:Geometry}}
\end{figure}

Fig.~\ref{fig:Geometry} displays both the experimental 3D hopper (a) as well as its simulation counterpart (b). The initial state of the hopper contains 50~\% more beads than shown in Fig.~\ref{fig:Geometry}(a). The red dye at the top was injected on top of the granular pile before the experiment and it propagates through the hopper faster than the grains can exit the system. (See the supplementary videos 1 and 2 which illustrate this process.)

The geometry is the same in the simulations and experiments, with very few exceptions. The initial filling height is smaller in the simulations, 
the hopper walls possess no thickness and have the same friction coefficient as the grains. The CFD simulation domain is divided into 1.5 million cells. The grid size gradually decreases near the hopper boundaries to ensure the quality of the solution in those areas. The meshing is realized applying the snappyHexMesh-tool embedded in the OpenFOAM software~\cite{openfoamdoc}.

In the simulations, the hopper flow is generated by first filling the hopper with the granular medium by pouring 
randomly the particles above the hopper top while the orifice remains closed. Then, once a sufficient filling height $h$ is
obtained, the granular packing is allowed to relax without the fluid for 0.5~seconds. At this point, the selection between the 
vacuum and submerged cases is made. In the vacuum case the orifice is opened, and the simulation is continued.
In the submerged case, the coupled CFD-DEM simulation is initiated and the orifice is opened.

\section{Results}

Motivated by the large computational cost of the prescribed numerical simulations, we revisit our earlier experimental findings with a new perspective.
The goal is to find a good compromise between having a long enough experiment with good surge per noise ratio (improves by reducing the particle size) and the computational burden (decreases with increasing particle size). The total particle number that can be handled with reasonable computational cost can be reached using the  average grain diameter of $d=0.2$~cm. 
Our main concern is the impact of the particle size on the surge. 
Hence, we start by comparing the earlier studied systems having $d=0.05$~cm and $d=0.1$~cm to the new system with $d=0.2$~cm. For this purpose, we observe the flow rate and compute selected dimensionless numbers characterizing the systems

\begin{figure}[!t]
\includegraphics[width=\columnwidth]{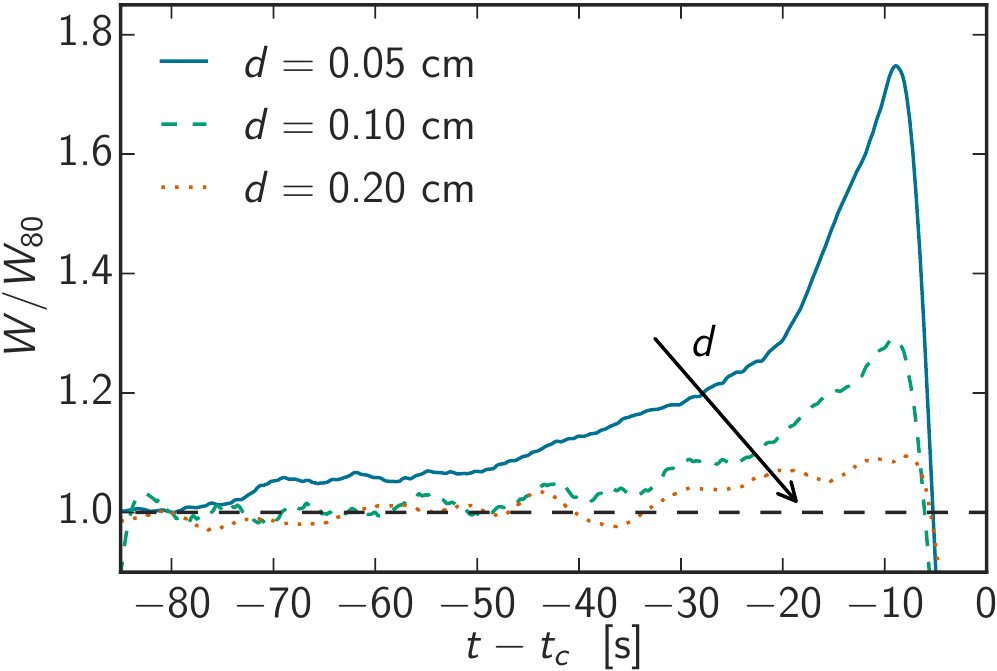}\\
 
\caption{The relative mass flow rate of hopper for three different particle sizes. The flow rate $W$ is normed with the flow rate $W_{80}$ that is the flow rate 80 seconds before the end. We find that the surge, the acceleration at the end decreases as particle size increases.
\label{fig:surge}}
\end{figure}
Fig.~\ref{fig:surge} shows the relative flow rate against time 
$t-t_{c}$, where the $t_{c}$ is the time when the flow stops. 
The flow rate $W$ is obtained by differentiating the mass time series of the 
scale by fitting a $2^{\mathrm{nd}}$ degree polynomial in a 2 second 
Gaussian window similarly to Ref.~\cite{koivistoSubmitted}.
As we are interested on the surge and dynamic effects we scale the data by the flow rate at $t=t_c-80$ s.
This operation allows us to compare the surge between the systems having different particle sizes.

The surge, the increase of flow rate $W$ with respect to the asymptotic value 
$W_{go}$ decreases with increasing particle size as highlighted by the black arrow in Fig.~\ref{fig:surge}. 
The largest increase $W_{surge} = \mathrm{max}(W) - W_{go}$ is with 
$d=0.05$~cm particles and the smallest is with $d=0.2$~cm particles. 
This agrees with earlier findings as the surge term 
containing the fluid-grain coupling has $d$ dependence as $W\propto(D-kd)^2$ 
(after expanding $\alpha$ from the supplementary material in 
Ref.~\cite{koivistoSubmitted}). 
The surge $W_{surge}$ thus decreases when approaching the clogging region 
from below by increasing the particle size. 
There is no flow, nor surge above the clogging region. 
We conclude that the large particles in our case approach to a limit where 
the granular aspect of the system starts to dominate. 
The inertia of the grains is too high for the fluid that there would be a 
large surge. 

Here, we would like to point out, that the superficial fluid velocity is faster than the grain velocity \cite{koivistoSubmitted}.
The fluid is faster and the inertia of the particles decreases the flow rate while the viscous component increases the flow.
This counterintuitive result is consistent with the earlier results 
\cite{koivistoSubmitted} and illustrated in supplementary video 1 with a 
layer of dye that propagates faster than the grains can exit. 
With small particles the fluid flow dominates the process and particles reflect to this.
The large particles have more momenta and inertia.
The fluid flow cannot affect the particle motion.
The granular characteristics of the large particles dominate.

\begin{figure}[!t]
\includegraphics[width=0.7\columnwidth]{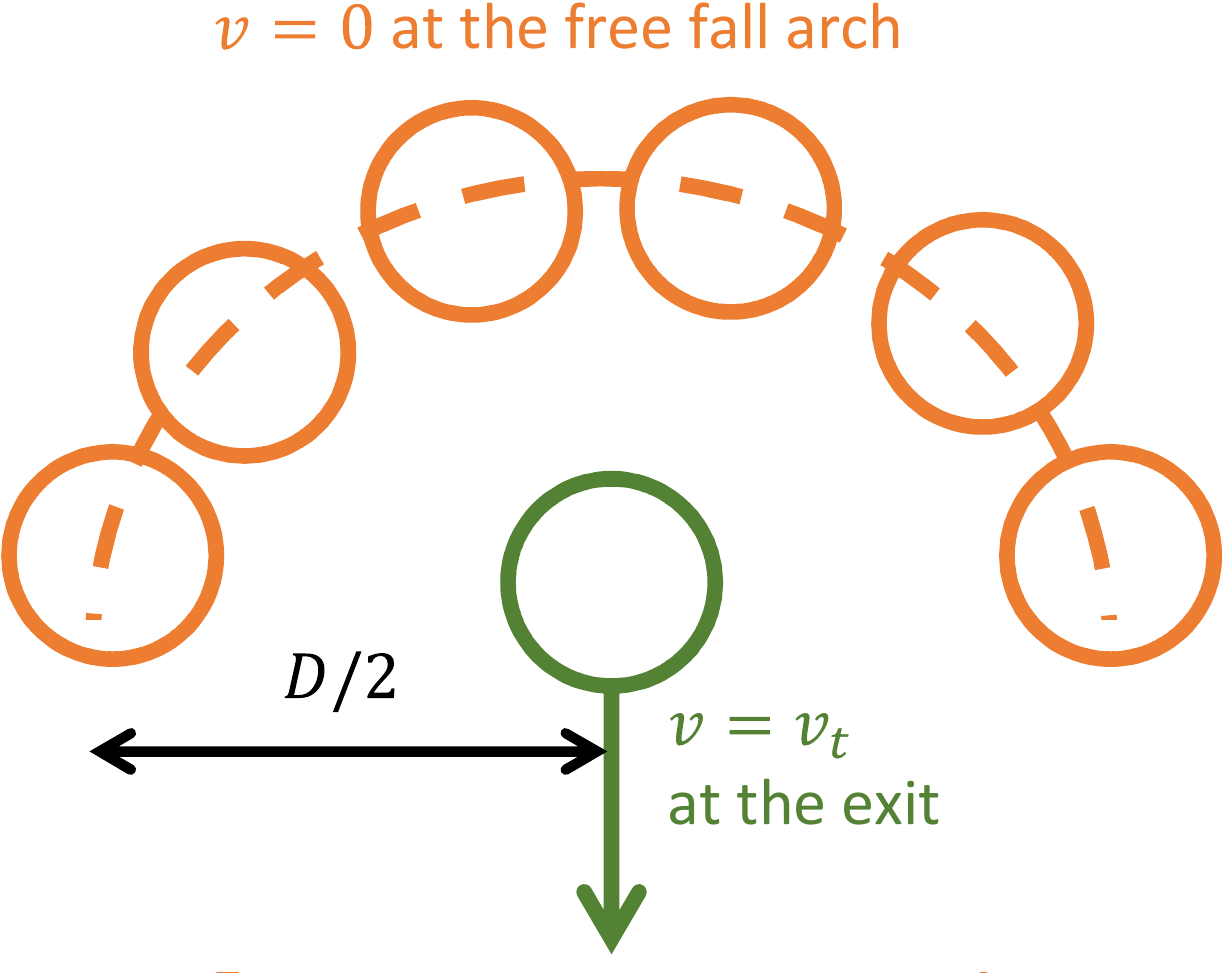}\\
 
\caption{Schematic illustration of the particles at the orfice of size $D$ accelerating from rest (orange) to terminal velocity (green) due to gravity and thus creating a shear rate and pressure.
\label{fig:freefall}}
\end{figure}

To obtain more rigorous treatment we calculate dimensionless numbers that describe the flow.
Table \ref{tab:dimensionless} describes the dimensionless numbers of the system.
The Reynolds number $Re = \rho_f v_t d/\eta$ describes the ratio of inertial forces respect to viscous forces. 
It increases dramatically as the particle diameter increases ($Re \approx d^2$) indicating
the increase of granular behavior at the expense of fluid flow, provided the grain properties (grain-grain friction, and grain size distribution) remain the same.
At the same time, the drag coefficient decreases~\cite{Morrison2013Drag}, again, indicating 
the diminishing effect of fluid. Note that here we discuss only the laminar flow case.
Finally we calculate the inertial number that is the ratio of confining 
pressure and shear rate $I=\eta \dot{\gamma}/P$ \cite{Houssais15NCO}. 
The inertial number $I$ can be approximated by defining the shear rate 
$\dot{\gamma} = v_t/(D/2)$ as velocity difference at the orifice and an 
approximation of driving pressure $P = 1/2\;\rho_{e} v_t^2$
as
\begin{equation}
 I 	= \eta \frac{\dot{\gamma}}{P} = \eta \frac{D}{\rho_{e}v_t}, 
\end{equation}
\noindent where the effective density $\rho_{e}$ is the buoyancy 
corrected density. 
The particle geometry at the orifice is illustrated in Fig.~\ref{fig:freefall}.
For small particles the inertial number is large, at the region where the 
dynamic effects already play a role. For large particles 
the inertial number decreases and the dynamic friction coefficient 
saturates (close) to static value \cite{Singh15NJP}.
This is seen as a lack of terminal surge as a constant dynamic friction
coefficient $\mu(I)$ indicates constant flow rate. 
Also, recently \cite{Trulsson16PRE} it has been numerically found that the ratio of frictional and viscous dissipation changes in submerged particle systems. Here, we are approaching the frictional regime from viscous regime by increasing the particle size leading to vanishing surge. 

\begin{table}[!h]
\caption{Dismensionless numbers describing the system. All numbers 
indicate that the effect of fluid as the dynamic viscous component is 
decreasing.}\label{tab:dimensionless}
\begin{tabular}{cccrr}
$d~~\mathrm{[cm]}$~	& ~~~$Re$~~~ 	& ~~~$C_d$~~~	&\multicolumn{1}{c}{~$I$~ }\\
\hline
0.05				& \Tstrut 37						& 1.76		&~$34.0 \times 10^{-4}$~\\
0.10				& 151						& 0.86		&~$17.0 \times 10^{-4}$~\\
0.20				& 530						& 0.56  	&~$ 9.8 \times 10^{-4}$~\\
\end{tabular}
\end{table}

 

The experimental study extends the research to larger particles in order
to reduce the particle number to a sufficient level to enable numerical 
simulations. 
The effect of grain size to the hopper flow is depicted in 
Fig.~\ref{fig:surge}, which shows the remaining mass of grains in the hopper
and the mass flow rate, both as a function of time for two different 
particle diameters. 
Not only the flow rate, but also the surge at the end of the experiment, 
depends on the particle diameter. As we have a grasp of the experimental 
aspects of the particle size dependence of the surge, it is possible to 
pick the largest particle size $d=0.2$~cm as a representative case.
 
\begin{figure}[!t] 
 \includegraphics[width=\columnwidth]{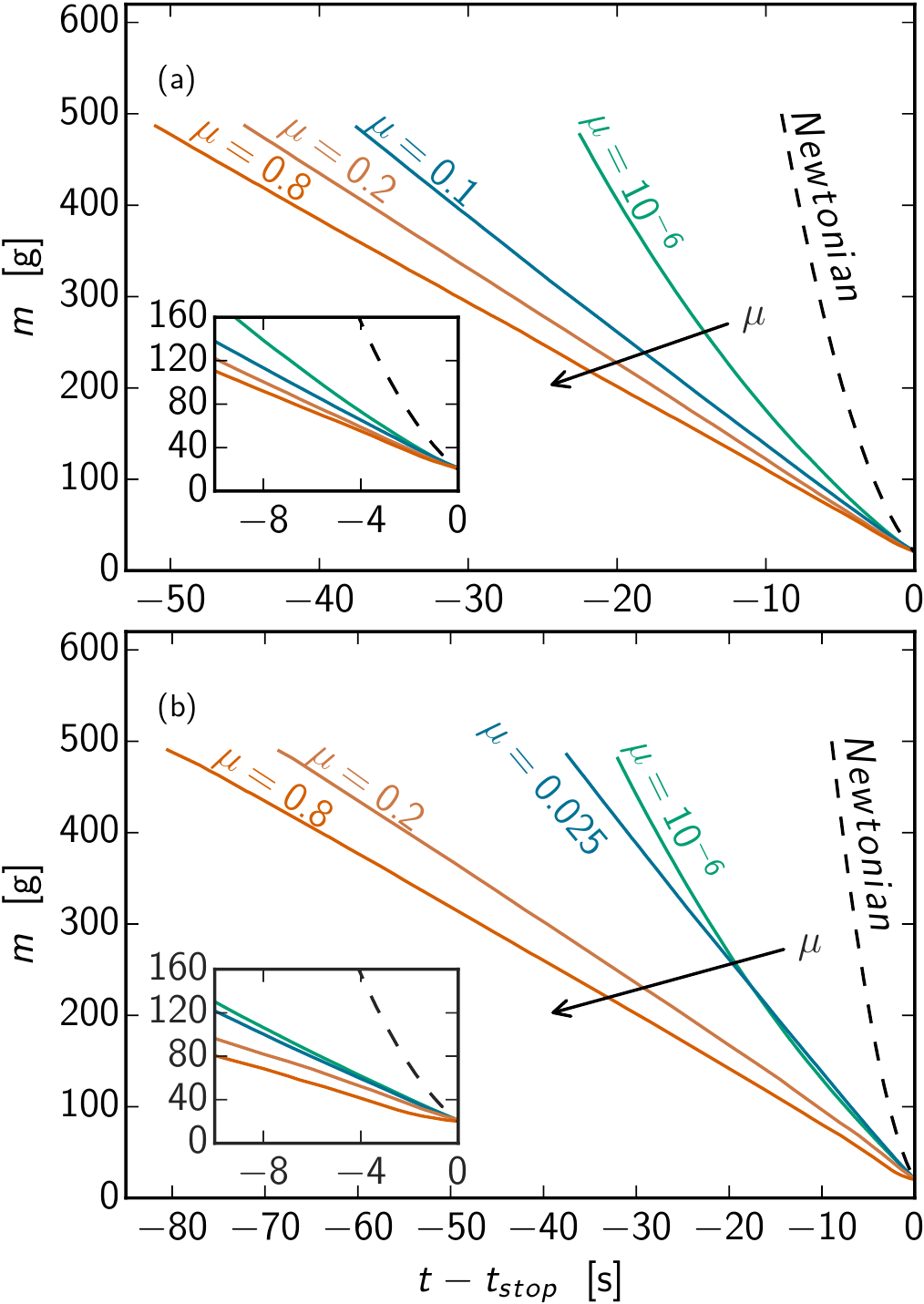}
\caption{(a) The mass of grains inside the hopper as a function of time for various values of friction coefficient $\mu$. For nearly frictionless case (green, $\mu=10^{-6}$) the flow rate (slope) is high and decreases as the grains flow out. This corresponds to what happens with a Newtonian fluid (dashed curve). For high friction case (red, $\mu=0.8$) the behavior is linear with constant flow rate and corresponds to the standard Beverloo case. (b) The mass of particles inside a hopper as a function of time similarly to Fig.~(a). For low friction coefficients the flow rate (slope) is decreasing similarly to the dry case. The insets show the magnification of the datasets near the end. }
 \label{fig:friction_dry} \label{fig:friction_sub}
\end{figure}
In the simulations the low friction granular (cyan) and the Newtonian fluid (dashed black) cases in Fig.~\ref{fig:friction_dry}(a) are non-linear and therefore not described by the Beverloo equation (\ref{eq:modified-beverloo}). 
The inset in Fig.~\ref{fig:friction_dry}(a) shows the magnification of the data near the end of the experiment. This is to point out that there seems to be no acceleration in the flow rate. 

Next, we repeat the simulation with parameters identical to the dry case with the exception that the grain-liquid coupling is enabled. Fig.~\ref{fig:friction_sub}(b) shows the mass in the hopper over $t-t_{c}$ as displayed earlier for the dry case in Fig.~\ref{fig:friction_dry}(a). The initial conditions, material parameters (except the friction coefficients), geometry, number of particles and even the initial particle locations are the same in each case. The largest difference is that the flow rates are significantly lower in the submerged cases. For instance, the simulation with the friction coefficient $\mu=0.8$ takes 52 seconds to empty 500 grams of grains in the dry case, while in the submerged case it takes 90 seconds. 

Additionally, there is an acceleration of the flow rate at the end of the simulation (Fig.~\ref{fig:friction_sub} insets). This is seen as separation of datasets and a slight downwards tilt in the data for the larger friction coefficients. Again, following the dry case, the low friction cases behave like Newtonian fluids without the acceleration. The contact friction of bulk granular materials is typically above $\mu=0.1$. For these values, we find a surge like feature in the submerged simulation, lacking from the dry case. As the only difference between the dry and submerged simulation is the inclusion of fluid, we conclude that the surge is due to the coupling between the liquid and grains.

\begin{figure}
\includegraphics[width=\columnwidth]{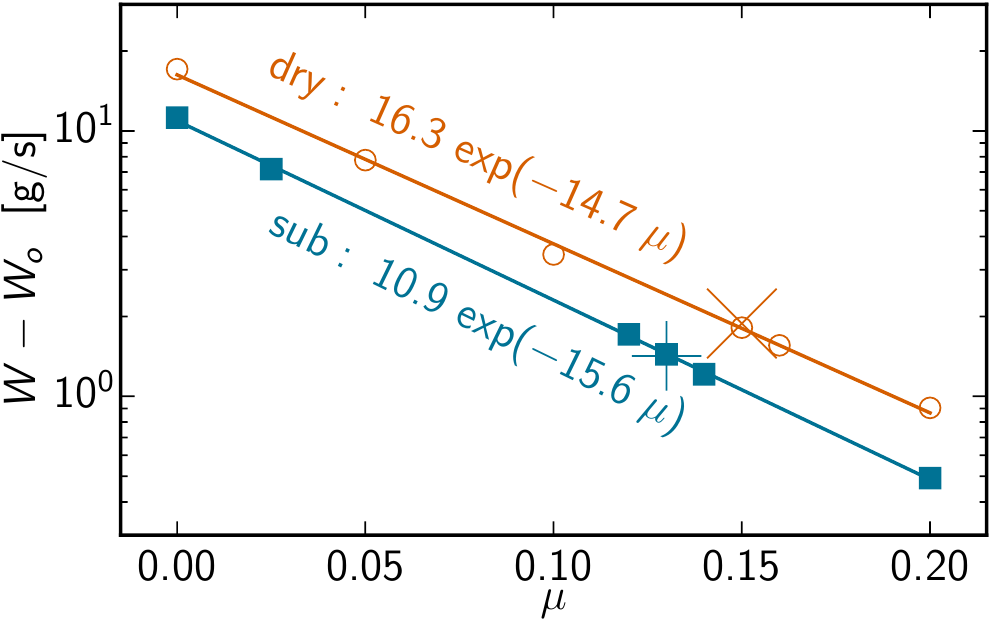}
\caption{Flow rate at the beginning of the experiment versus friction coefficient for dry (red open circles) and submerged (blue filled squares) extracted from previous figures. The fit and functional form is purely empirical and used in finding the matching friction coefficient. The value of $W_{o,dry} = 9.5~\mathrm{g/s}$ and $W_{o,sub} = 6.1~\mathrm{g/s}$. The behavior appears to exponential in this narrow region. Here we anticipate the experimental results plotting the mass flow rate in dry experiments (red x) at $W_{dry} = 11.4~\mathrm{g/s}$ corresponding friction coefficient $\mu_{dry}=0.15$ in the simulations. Similarly, the mass flow rate in submerged experiments (blue +) is $W_{sub}=7.6~\mathrm{g/s}$ corresponding to friction coefficient $\mu_{sub} = 0.13$ in simulations.}\label{fig:q_vs_mu}
\end{figure}

In Fig.~\ref{fig:q_vs_mu}, we plot the simulated reduced flow rate $W-W_o$ with multiple values of the friction coefficient $\mu$, creating an empirical relation between the initial flow rate $W$ and the friction coefficient $\mu$. Here, $W_o$ is one of the fitting parameters. Based on this empirical relation we deduce the friction coefficient by matching the flow rates in the experiments at $m=300\ldots 400$~g. The red open circles correspond to the dry simulations  and the filled blue squares to the submerged simulations. The friction coefficients that reproduce the experiment are almost equal as $\mu_{dry}=0.15$ and $\mu_{sub}=0.13$ in the dry and submerged cases, respectively. The similarity in dry and submerged friction coefficient is also reported by Dijksman {\it et al.}~with acrylic beads in a rheometer \cite{DijksmanPRE10}. Note that here we refer to grain-grain friction $\mu$ whereas Dijksman {\it et al.}~refers to the  minimum friction coefficient $\mu_o$ at the quasi-static limit when inertial effects vanish $I\to 0$. The relation $\mu_o(\mu)$ is a non-trivial monotonic function that (to our knowledge) is only explored numerically \cite{Lemaitre09RHA,DaCruz05PRE,Trulsson16PRE} 

%



\begin{figure}[!t]
 \includegraphics[width=\columnwidth]{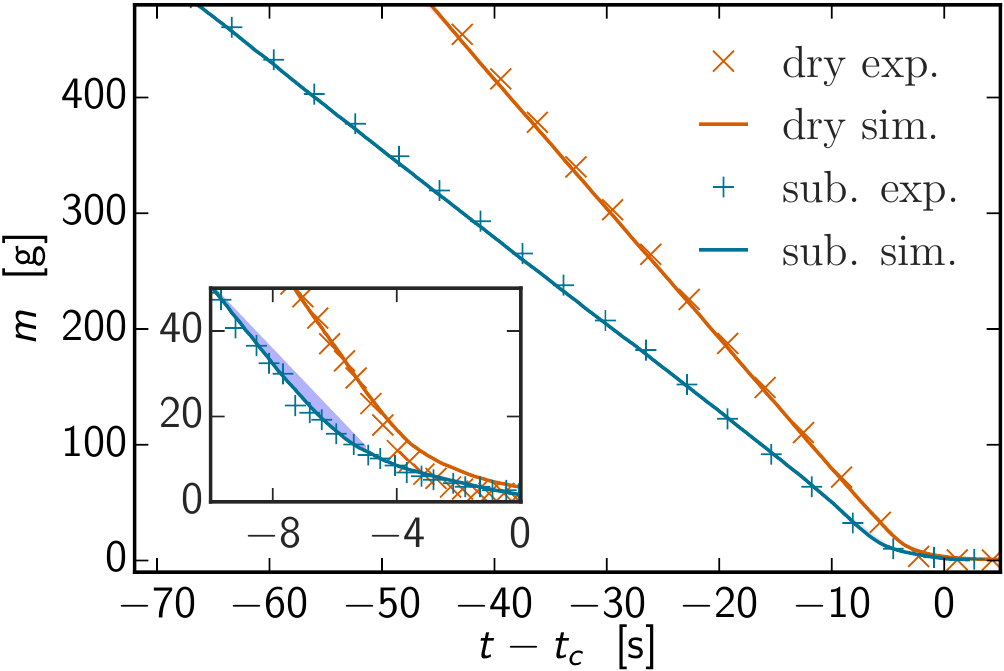}\\
 \caption{The raw data from the experiments and simulations show the difference between dry (red) and submerged (blue) case. The Linear fit depicted with solid line to the simulation data above $m = 200$~g matches the dry case perfectly until the grains run out. In submerged case the data takes a nose dive, surges, before the grains run out. Inset: the blue swath indicates a surge, difference between linear Beverloo behavior and measured data.}
 \label{fig:flow_rates}
\end{figure}

Fig.~\ref{fig:flow_rates} displays the mass remaining in the hopper from both the experiments and the simulations with the friction coefficient set to the obtained values of $\mu_{dry}=0.15$ in the dry case and $\mu_{sub}=0.13$ in the submerged case. The submerged experiment is depicted in blue and the dry case in red color. The datasets are the result of a single run. The simulated and the experimental data are overlapping within the measurement accuracy. This lends credence to the computational approach applied in the work and specifically suggests that the coupled CFD-DEM model captures the quintessential features of the two-phase (submerged) hopper flow.
%

\begin{figure}[!ht]
 \includegraphics[width=\columnwidth]{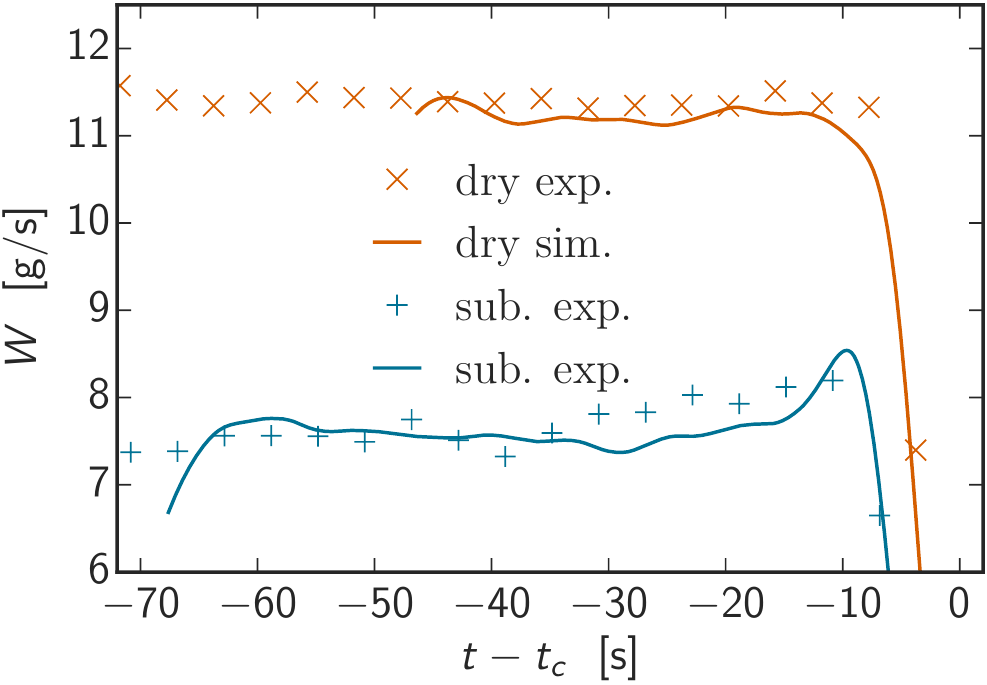}\\
 \caption{The derivative of hopper mass over time shows the surge in submerged case for both experiments (blue +) and simulations (solid curve). The surge is not seen in the dry experiments (red x) and simulations (solid red). However, the final moments of the dry experiment might contain a tiny surge that is too fast for the current experimental procedure and analysis.}
 \label{fig:derivated_data}
\end{figure}

Fig.~\ref{fig:derivated_data} displays the Gaussian weighted derivative over two second time window of the data depicted in Fig.~\ref{fig:flow_rates}. In both dry cases, the experiment and the simulation, the hopper empties at a constant flow rate. 
At the end when the grains run out and the flow rate decreases without a terminal surge. 
In contrast, the presence of the interstitial fluid reduces the overall granular flow rate and imposes an acceleration towards the end. Recently, such terminal surge has been confirmed in the dry case for smaller particles in experiments \cite{koivistoSubmitted} and appears to be visible also in simulations \cite{DunatungaJFM15,SchwartzGM12}.

Based on our theoretical discussion on the terminal velocity, and on the dependence of the surge on particle size, we propose that the 2~mm particles are too heavy to be affected by interstitial air. Therefore, the surge does not appear in the dry experiments resulting in good agreement to our vacuum simulations. Since the viscosity and density of water are several orders of magnitude larger, the submerged flow exhibits a surge. We conclude that the viscosity of the interstitial medium has to be large enough compared to the particle inertia for the surge to appear.
%

\section{Conclusions}

We performed experiments and simulations on dry and submerged hopper 
flows of granular particles of approximately millimeter radius using the 
combination of DEM and CFD. 
In the frictionless case, we confirm the previously known numerical result 
\cite{Langston1994}, that the flow rate of the dry granular particles 
decreases as a function of time. 
Here, we have extended the result of decreasing flow rate to submerged 
conditions.
This scenario, which occurs in both dry and submerged system could be 
understood in the context of a Newtonian fluid running out of a hopper.

The grain-grain friction changes the scenario from a Newtonian behavior into a more complex one: In the dry case the grain flow rate remains constant until the height of the granular column is less than the width of the hopper \cite{Nedderman1982}. This is a well understood behavior, and is readily described by the Beverloo equation.

\begin{figure}[!t] 
 \includegraphics[width=\columnwidth]{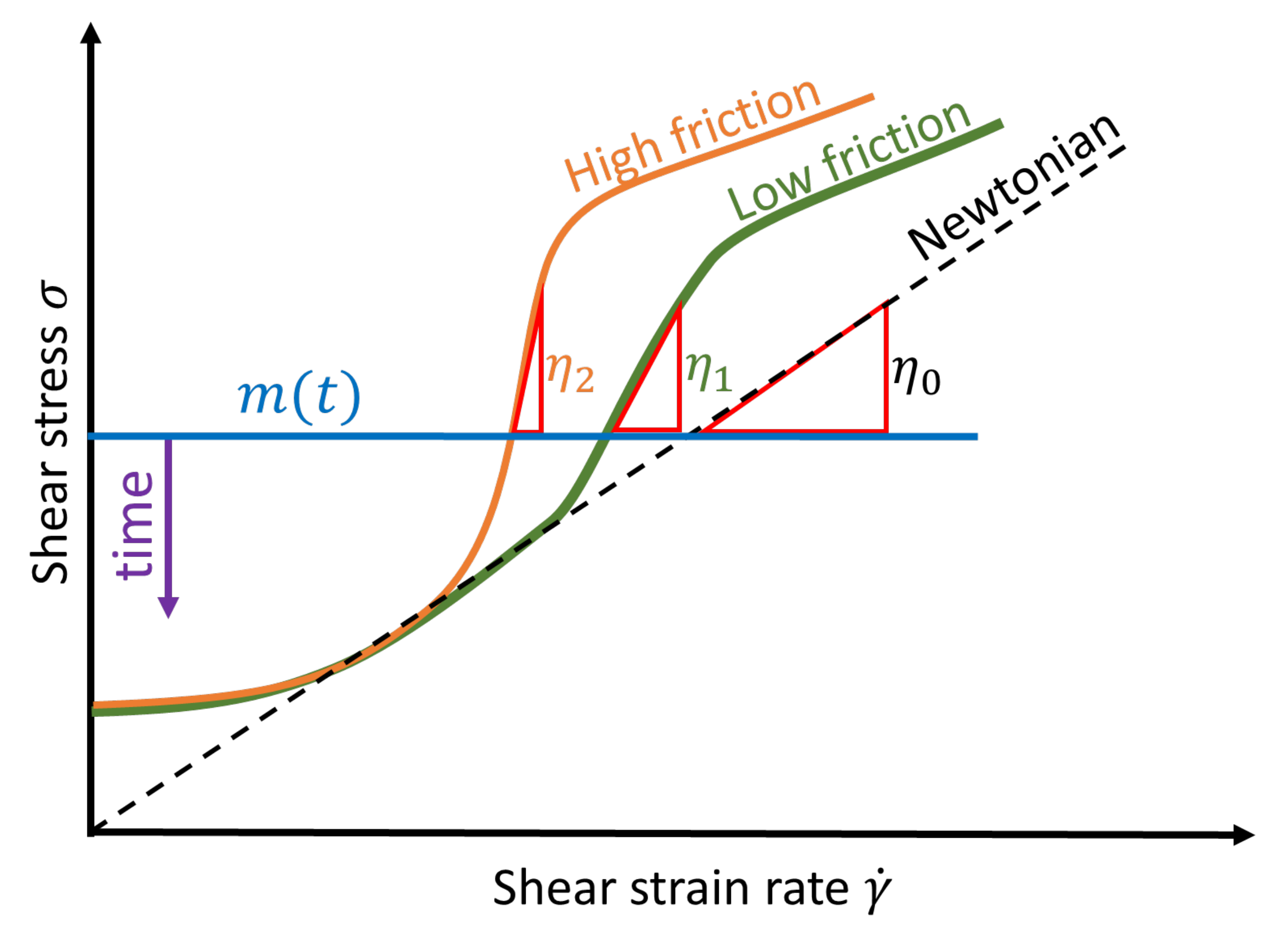}\\
 \caption{A schematic illustration of three materials with discontinuous shear thickening. The increasing contact friction of the particles leads to decreasing flow rate. At continuum, this can be interpreted as increasing effective viscosity $\eta_{i}$ that increases with particle friction for high shear stresses.}\label{fig:DST}
\end{figure}

In the submerged hopper case, the grain-grain friction has an even more profound effect. The 
flow accelerates, surging through the whole hopper emptying process. Furthermore, 
right before the hopper runs out of grains there is a clear terminal surge in the flow rate. The
accelerating flow can be understood in the framework of a feedback loop mediated by the
incompressible, viscous water. The grains exit the hopper as the gravity pulls them. 
The grains replace a certain volume of water outside the hopper. Due to 
the water incompressibility the same amount of water needs to enter the hopper. This mainly
occurs through the open top, where the flow resistance is the smallest. That creates a flow of
water through the granular packing, which in turn mediated by the viscosity, increases the outflow of the
grains. This granular pumping effect is described in \cite{koivistoSubmitted} and is captured by the 
simulation here. 

As we observe, both the dry and submerged cases are sensitive to the grain-grain friction.
This allows us to use the parameter to fit the simulated flow rates against the corresponding 
experiments. Subsequently, we observe that the optimal friction parameter is almost
the same in both the cases. This was a surprising result, since one would expect the grain-grain
friction to be significantly lower on a wet surface. However, as we do not explicitly account for the grain-grain hydrodynamic interactions, we expect the
friction coefficient to partly compensate for that. This sensitivity to friction coefficient gives
the possibility to interpret the hopper flow in the context of non-linear effective rheology.

Fig.~\ref{fig:DST} shows a schematic illustration of three systems with (discontinuous) shear thickening \cite{Peters2016}, a characteristic of frictional granular systems. For the same load, caused by the high particle column, the frictionless case, a Newtonian fluid, has the smallest slope and thus lowest effective viscosity. Friction increases the slope and introduces a sudden increase of viscosity, that can be many orders of magnitude. This is observed here as rapid decrease of flow rate as a function of particle-particle friction.
When the mass $m(t)$ of the particle column decreases in time, the effective viscosity of the system decreases as well, causing an increase in the flow rate. This is seen as the terminal surge. For a system of high interparticle friction this has a greater impact.

Here we have presented the first step to compare simulations to the experiments with a good 
agreement. The one-to-one match with experiments and simulation is currently pushing the 
limits of both methods.
Using smaller than $d=0.2$~cm particles increases the experimental accuracy via lowering the flow rate. 
However using smaller particles renders the simulation impractical by making the problem too large for the present computational resources. Both of these problems, the experimental and numerical are solvable in the near future by advanced computational methods and measurement techniques. 
Future studies should involve 
the effect of wall and bottom friction, dilation of grains at the orifice, clogging, self-generated pumping of fluid, terminal and exit velocities of particles. 
Also, we would like to point out that the behavior of the flow rate at infinitely 
tall packings, the constant component $W_{go}$  as the function of particle size 
is also interesting. There is a transition from colloidal no-flow behavior to surging flow and back to no-flow 
at clogging, but this is outside the scope of this paper.

\section{Acknowledgement}

This work was supported by the Academy of Finland
through the COMP center of excellence and the project
number 278367. The simulations were performed using
the computer resources within the Aalto University
School of Science, Science-IT project. JK acknowledges 
the support from Finnish cultural foundation and Wihuri 
foundation through Foundations' post doc pool project. DJD acknowledges 
the support from NFS through grant DMR-1305199.

\addcontentsline{toc}{section}{References}

\end{document}